\documentstyle[prb,aps,multicol,epsfig]{revtex}

\draft
\begin{document}
\title{Surface induced ordering in thin film diblock copolymers:
tilted lamellar phases}
\author{Y. Tsori\footnote{email: tsori@post.tau.ac.il} 
and D. Andelman\footnote{email: andelman@post.tau.ac.il}}
\address{School of Physics and Astronomy,
Raymond and Beverly Sackler Faculty of Exact Sciences\\
Tel Aviv University, 69978 Ramat Aviv, Israel}
\date{3/6/2001}

\maketitle


\baselineskip =16pt  
\begin{abstract}
We investigate the effect of chemically patterned surfaces on the
morphology of diblock copolymers below the order-disorder
transition. Profiles for lamellar phases in contact with one
surface, or confined between two surfaces are obtained in the
weak segregation limit using a Ginzburg-Landau expansion
of the free energy, and treating it with mean-field theory. The
 periodically patterned surface induces a
tilt of the lamellae in order to match the surface periodicity.
The lamellae relax from the constrained periodicity close 
to the surface to the bulk periodicity far from it.
The phases we investigate are a generalization to the mixed
(perpendicular and parallel to the surface) lamellar phases
occurring when the two surfaces are homogeneous. A special case
when the surface pattern has a period equal to the bulk lamellar
period showing ``T-junction'' morphology is examined. Our analytic
calculation agrees with previous computer simulations and self
consistent field theories.

\end{abstract}


\baselineskip=12pt 

\begin{multicols}{2}
\section{Introduction}\label{intro}

 Diblock copolymers are made up of two chemically
distinct polymer chains which are covalently bonded together. The
process of macroscopic phase separation usually occurring for
incompatible chains is suppressed because of the imposed
connectivity between the two blocks. Instead, the system may
undergo a mesoscopic phase separation. The thermodynamic state of
the system depends on two parameters \cite{B-F90,O-K86,Leibler80,F-H87,M-B96}:
the fraction of the A monomers $f=N_A/N$ and $\chi N$,
 where $N_A$ and $N_B$ are the
lengths of the A- and B-block, respectively, and $N$ is the
polymerization index, $N=N_A+N_B$. The
 Flory parameter
$\chi$ measures the interaction energy (in units of $k_BT$, the
thermal energy) between two monomers and is positive if the two
monomers repel each other.
 For symmetric melt ($f=1/2$)
and small $\chi$ (high temperature) the disordered, homogeneous
phase has the lowest free energy. Increasing $\chi$ above the order
disorder transition (ODT) point, $\chi=\chi_c$, (lowering the
temperature) results in a phase transition to a lamellar phase. For
a fixed $\chi>\chi_c$, and as function of the A/B asymmetry ($f\ne
1/2$)  the prevailing mesoscopic ordered phases can have hexagonal
or cubic symmetries as well. \cite{Leibler80,F-H87}

When a block copolymer (BCP) melt is put in contact with a
surface, the surface reduces the chain entropy. In addition, it
interacts chemically with the blocks, leading to a surface behavior
which can be very different from that of the
bulk. Fredrickson \cite{Fredrickson87} has considered BCP in
contact with a uniform surface having preferred interaction to one
 of the blocks. In the weak-segregation regime, he used mean-field
 theory
to investigate systems both below and above the ODT. Above, but
close to the ODT, the order parameter (being the deviation of the A
monomer concentration from its average value $f$) has decaying
oscillations characterized by a correlation length $\xi$.
Approaching the ODT, $\xi$ diverges and below the ODT the system is
characterized by a spatially modulated order parameter.
\cite{Fredrickson87}

The situation is more complex for a copolymer melt confined between
two surfaces.  For a thin-film of BCP melt, the interplay of the
spacing $2L$ between the surfaces, the Flory parameter
$\chi$ and the surface interactions results in a rich interfacial
behavior. \cite{G-M-B00,epl01}
The phase behavior of thin BCP films subject to uniform surface
 fields has been
investigated numerically using self-consistent fields (SCF) theory
\cite{matsenJCP97,pbmm97} and Monte-Carlo simulations,\cite{G-M-B00,wang1} and
was found to consist of parallel, perpendicular and mixed lamellae
denoted $L_\parallel$, L$_\perp$ and $L_M$, respectively. The latter
$L_M$ phase has parallel lamellae extending from one surface, which
are jointed in a
 T-junction defect with perpendicular lamellae extending from the
opposite surface.\cite{P-WMM99,wang2}
At a given inter-surface spacing, increasing the
(uniform) surface interactions promotes a parallel orientation
with either A-type or B-type monomers adsorbed to the surface.
However, if the spacing $2L$ between the surfaces is
incommensurate with the lamellar periodicity, or the incompatibility
$\chi$ is increased, a perpendicular orientation is favored.
 \cite{T-Axx00}

In the present paper we analytically derive expressions for the
order parameter of a BCP melt confined between two parallel flat
surfaces, below the bulk ODT temperature and in the
weak-segregation limit. In Sec. \ref{model} we introduce a model of
a confined BCP melt in contact with either homogeneous or
sinusoidally patterned surfaces.\cite{P-WMM98}
Contrary to the system above the
ODT, studied earlier in a separate work,\cite{epl01,mm01} a linear response
theory assuming small order parameter as a response to the surface
fields is inadequate, since the bulk phase has an inherent spatially
varying structure. Instead,  in Sec. \ref{1surface} an expansion is
carried out around a tilted lamellar phase of the bulk. 
A simple ansatz for the
deviation of the order parameter from its bulk value is suggested,
utilizing the symmetry of the problem. A close similarity is shown
between this problem and the symmetric tilt grain boundaries, with
mathematical solution very similar to what has been derived in
Ref.~\onlinecite{T-A-S00}.  In Sec. \ref{2surfaces} the problem of
one patterned surface is extended to a melt confined between two
surfaces, one being chemically patterned while the second is
chemically homogeneous. We obtain confined tilted lamellar phases, and, in
particular, $L_M$ phases. The lamellae extending from the patterned
surface merge with the parallel ordering induced by the
homogeneous surface.

\section{The model}\label{model}

When a BCP system in the disordered (high temperature) phase is
cooled below the ODT, a lamellar phase forms
\cite{N-A-SPRL97,V-N-A-S98} for $f=1/2$ (symmetric) or when
$|f-1/2|$ is small.
The behavior of such systems was modeled
numerous times in the past
\cite{F-H87,epl01,mm01,binder97,SH77,C-C98}
using 
coarse grained Ginzburg-Landau free energy. A possible
form (in units of $k_B T$) used throughout this paper is: 
\begin{equation}  \label{Fb}
{\cal F}_b=\int\left\{\frac12\tau\phi^2+\frac12h\left[\left(\nabla^2+
q_0^2\right)\phi\right]^2 +\frac{u}{4!}\phi^4-\mu\phi
\right\}{\rm d}^3{\bf r}
\end{equation}
where
\begin{eqnarray}
q_0\simeq 1.95/R_g,\\
\tau=2\rho N\left(\chi_c-\chi\right),\\
\chi_c\simeq10.49/N,\\
h=3\rho c^2 R_g^2/2q_0^2.
\end{eqnarray}
The copolymer order parameter $\phi$ is
 defined as $\phi({\bf r})\equiv\phi_A({\bf r})-f$, the deviation
  of the local A monomer concentration from its average value.
$R_g$ is the gyration radius of the chains ($R_g^2\simeq
\frac16Na^2$ for Gaussian chains) and the chain density $\rho$ is
equal to $1/Na^3$ for an incompressible melt. The chemical
potential is $\mu$ while $u/\rho$ and $c$ are dimensionless
parameters of order unity. Hereafter we set the monomer size to
unity, $a=1$, expressing all lengths in units of $a$. By choosing
$u/\rho=c=1$ (and recalling that $f=1/2$), all parameters in the
model are given in terms of $\chi$ and $N$.

Above the ODT, $\chi<\chi_c$, the free energy (\ref{Fb}) describes a
system in the homogeneous, disordered phase having a uniform order
parameter $\phi=0$. This system has been previously studied by us
in Ref.~\onlinecite{mm01}. Below the ODT, $\chi>\chi_c$, the
lamellar phase of period $d_0=2\pi/q_0$ becomes stable. Close to
the ODT, $\chi\gtrsim\chi_c$, (weak segregation limit), the order
parameter is given in the single $q$-mode approximation as
$\phi({\bf r})=\phi_0+\phi_q\cos({\bf q_0\cdot r})$, and for
symmetric BCP ($f=1/2$), used throughout the paper,
$\langle\phi({\bf r})\rangle=\phi_0=0$. It is worthwhile to mention
that similar free energy functionals have been used to describe
bulk and surface phenomena in diblock
copolymers,\cite{Leibler80,F-H87,T-A-S00,N-A-SPRL97} amphiphilic
systems, \cite{G-S90} Langmuir films \cite{A-B-J87} and magnetic
(garnet) films.\cite{G-D82}

The interaction of the BCP with the confining surfaces is assumed
to be short-range and limited to the surface only. The surface free
energy ${\cal F}_s$ (in units of $k_BT$) has the form:
\begin{equation}\label{Fs}
{\cal F}_s=\int\left\{\sigma({\bf r_s})
\phi({\bf r_s})+\tau_s\phi^2({\bf r_s})\right\}{\rm d^2{\bf r_s}}
\end{equation}
where the integration is carried out over all surface positions $\{{\bf
r_s}\}$. At a point ${\bf r_s}$ on the surface, the first term is
proportional to the order parameter, with $\sigma({\bf r_s})$ being
a surface field. The magnitude $\sigma({\bf r_s})$ of this surface
field can be controlled by coating the substrate with random
copolymers.\cite{L-RPRL96,M-RPRL97} The second (quadratic) term
allows us to describe cases where the local surface segregation is
different than the bulk: a positive $\tau_s$ means local increase
of the Flory parameter $\chi$ and a lower phase transition temperature.

The surface effects are
contained in the correction to the order parameter
\begin{equation}
\delta\phi({\bf r}) \equiv\phi({\bf r})-\phi_b({\bf r})
\end{equation}
where the bulk order parameter $\phi_b({\bf r})$ 
in the symmetric lamellar phase is given by
\begin{eqnarray}
\phi_b({\bf r})&=&\phi_q\cos\left({\bf q_0\cdot r}\right)
~~~,\\
\phi_q^2&=&-8\tau/u~~~.
\end{eqnarray}
The bulk $\phi_b$ does not depend on
surface properties, and recall that $\tau<0$ below the ODT.
The total free energy ${\cal F}={\cal F}_b+{\cal F}_s$ is now
expanded about its bulk value ${\cal F}[\phi_b]$ to second order in
$\delta \phi$ : ${\cal F}={\cal F}[\phi_b]+\Delta {\cal F}$, with

\begin{eqnarray}\label{deltaF}
\Delta {\cal F} &=&\int \left\{[(\tau+hq_0^4)\phi _b+\frac16u\phi
  _b^3+hq_0^2\nabla^2\phi_b-\mu]\delta \phi\right.\\
&+&\left.\frac12(\tau+\frac12u\phi_b^2)\left(\delta\phi\right)^2
+\frac12h\left[\left(\nabla^2+q_0^2\right)\delta\phi\right]^2
\right\}{\rm d}^3{\bf r}\nonumber\\
&+&\int \left\{\sigma({\bf r}_s) \delta\phi({\bf r}_s)+\tau_s\left(2\phi_b({\bf r}_s)\delta\phi({\bf r}_s)+
\delta\phi({\bf r}_s)^2\right)\right\}{\rm d}^2{\bf r}_s\nonumber
\end{eqnarray}
This expansion is valid in the weak segregation limit and 
for small enough surface fields.
In the next section we find the function $\delta\phi(x,y,z)$ that
minimizes the free energy functional $\Delta {\cal F}$ 
for a given choice of bulk morphology
$\phi_b({\bf r})$ and surface field $\sigma({\bf r_s})$.

\section{BCP melt confined by one patterned surface}\label{1surface}


When the confining substrate is spatially modulated, at least one
additional length scale enters the problem. Numerical studies of thin
BCP film below the ODT temperature in the presence of
sinusoidally patterned surfaces of period $d_x$ have been carried out by Petera
and Muthukumar.\cite{P-Muthu98} It was found that the lamellae are
tilted with an angle $\theta\equiv\arccos(d_0/d_x)$ with respect
to the normal to the surface.
We proceed along similar lines, assuming that the pattern located
at $y=0$ [see Fig.~1(a)] is described by a single harmonic,
\begin{equation}\label{sigma2}
\sigma(x,z)=\sigma(x)=\sigma_0+\sigma_q\cos(q_xx)
\end{equation}
and is translational invariant in the $z$-direction. The parameter
$\sigma_q$ sets the strength of the modulation mode, while the
average $\langle\sigma(x)\rangle$ is given by $\sigma_0$. At this
point we assume $\sigma_0=0$, hence the surface is overall neutral
to the A/B adsorption. Furthermore, the periodicity $d_x=2\pi/q_x$
along the $x$--axis is assumed to be longer than the natural
lamellar periodicity $d_0$ in the bulk, $d_x>d_0$.

The surface field $\delta\phi$ is the response of the BCP system
to the presence of the surface field $\sigma$. It is
expected to
vanish far away from the surface:
\begin{equation}\label{bcphi}
\lim_{y\rightarrow\infty}\delta\phi(x,y,z)=0
\end{equation}
recovering the bulk $\phi=\phi_b$ phase. In general, the bulk
lamellae can be tilted with respect to the $y=0$ surface:
\begin{equation}
\phi_b(x,y)=\phi_q\cos\left(q_0\cos\theta x +q_0\sin\theta y
+\alpha\right)
\end{equation}
and this form of $\phi_b$ will be used as a zeroth order
approximation to $\phi(x,y)$. Note that the bulk free energy Eq.
(\ref{Fb}) is invariant with respect to rotations and translations
of the lamellae, and does not depend on the tilt angle $\theta$ and
the phase shift $\alpha$. These parameters are chosen such that
they minimize the surface free energy Eq. (\ref{Fs}). The tilt
angle $\theta$ determines the overlap of the lamellae with the
surface inhomogeneities, while the phase shift $\alpha$
distinguishes between two identical tilts, in-phase ($\alpha=0$) and
out-of-phase ($\alpha=\pi$) with respect to the surface. Upon
integrating the $x$ variable, the surface free energy Eq.
(\ref{Fs}) is minimized if the angle $\theta$ obeys
$\cos\theta=q_x/q_0$ and $\alpha=\pi$.
Therefore, we use below a bulk phase $\phi_b$ given by
\begin{eqnarray}  \label{bulk}
\phi_b=-\phi_q\cos\left(q_xx+q_yy\right)\\
q_x=q_0\cos\theta,\qquad\qquad q_y=q_0\sin\theta,
\end{eqnarray}

For the correction order parameter $\delta\phi$ we choose
\begin{equation}\label{dphi}
\delta\phi(x,y)=g(y)\cos(q_xx).
\end{equation}
This correction describes a lamellar ordering perpendicular to the
surface, and commensurate with its periodicity $d_x=2\pi/q_x$. The
overall morphology of the lamellae is a superposition of the
correction field $\delta\phi$ with the bulk tilted phase, having a
periodicity $d_0$. The region where the commensurate correction
field $\delta\phi$ is important is dictated by the amplitude and
range of the amplitude function $g(y)$. It is now possible to use
ansatz (\ref{dphi}) to integrate out the $x$ dependence in the free
energy, Eq. (\ref{deltaF}), resulting in an effective
one-dimensional free energy \cite{T-A-S00}
\begin{eqnarray}\label{Dg}
\Delta F(y)&=&-\frac12\tau\cos^2(q_yy)g^2
+\frac14h\left(q_y^2g+g^{\prime\prime}\right)^2\nonumber\\
&+&\frac12 \left[\sigma_qg+\tau_s\left(-2\phi_q g+g^2\right)\right]\delta(y)
\end{eqnarray}
For convenience we have included the surface free energy terms of
Eq. (\ref{Fs}) by the use of a Dirac delta function $\delta(y)$.

Applying the variational principle with
respect to $g(y)$ yields a master equation:
\begin{equation}
\label{gov_geqn}
\left[ A+C\cos(2q_yy) \right]g(y)+Bg^{\prime\prime}(y)
+g^{\prime\prime\prime\prime}(y)=0,
\end{equation}
with parameters $A,B$ and $C$ given by:
\begin{eqnarray}
A&=&-\tau/h+q_y^4,\nonumber\\
B&=&2q_y^2,\nonumber\\
C&=&-\tau/h.
\end{eqnarray}
The equation is linear in $g(y)$ since the free energy is expanded
 to second order
around its bulk solution $\phi_b$. Clearly the bulk solution
($g\equiv0$) solves the Euler-Lagrange equation, and hence Eq.
(\ref{gov_geqn}) is homogeneous. The boundary conditions on the
$y=0$ surface are
\begin{eqnarray}\label{g_bcs}
\left(\sigma_q-2\tau_s\phi_q\right)/h+2\tau_sg(0)/h+
q_y^2g^{\prime}(0)+g^{\prime\prime\prime}(0)&=&0, \nonumber \\
q_y^2g(0)+g^{\prime\prime}(0)&=&0.
\end{eqnarray}
Hence the fourth order differential equation
(\ref{gov_geqn}) has to satisfy two boundary conditions, mixing the
value $g(0)$ and its first three derivatives at the surface.

 Consider the Euler-Lagrange equation (\ref{gov_geqn}) obeyed
by the function $g(y)$. It is similar to the Schr\"{o}edinger
equation for an electron in a periodic one--dimensional potential
$V(y)=-C\cos(2q_yy)$, the ``kinetic energy'' term being proportional
to $B$, and the electron `` total energy'' $A$. Note that unlike the
quantum mechanical problem, here we have in addition a fourth order
derivative term $g^{\prime\prime\prime\prime}$. Moreover, the
coefficient $B$ depends on the periodicity of the potential.
Nevertheless, the general form of $g(y)$ can be written, similar to
the quantum mechanical problem, in terms of the Bloch form:
\begin{equation}\label{bloch}
g(y)=e^{-ky}\sum_na_ne^{2inq_yy}\quad+\quad c.c.
\end{equation}

The surface induced deformations of lamellar phases are similar to
the deformations appearing in the symmetric tilt grain boundaries.
\cite{G-TMAC94} The plane of symmetry between adjacent grains
corresponds to the solid confining surface of our system, and in
both cases packing frustration near the interface plays an
important role. However, the fact that the surface can have
(nonuniform) interactions with the copolymer melt results in a
richer set of phenomena in the system studied here.

We now briefly mention how Eq. (\ref{gov_geqn}) can be solved. More
details can be found in Refs.~\onlinecite{T-A-S00} and
\onlinecite{M-F}. The form (\ref{bloch}) is substituted in Eq.
(\ref{gov_geqn}) yielding a recurrence relation between the
coefficients $\{a_n\}$. 
%
%
Only for a special value of the eigenvalue (wavenumber) $k$,
the series in Eq. (\ref{bloch}) converges. Once $k$ is determined iteratively,
the full solution of Eq. (\ref{gov_geqn}) is given analytically by
Eq. (\ref{bloch}).
 If $k$ is a solution, then so are $-k$, $k^*$ and
$-k^*$, recalling that Eq. (\ref{gov_geqn}) is a fourth-order
differential equation. The periodicity of the ``potential term'' is
determined by $q_y$, which depends on the surface periodicity
 $d_x=2\pi\tan\theta/q_y$.
Modification of $d_x$ smoothly changes ${\rm
 Im}(k)$ until it becomes equal to an integer multiple of $q_y$.
   At
these degenerate values of $d_x$, the plot
of ${\rm Re}(k)$ as a function of $d_x$  has two branches, each
 branch
corresponding to one possible solution of $k$.
The values of $d_x$
where the degeneracy appears are determined by the proximity to the
ODT.
For a semi-infinite BCP bounded by a single surface, the function
$g(y)$ vanishes at infinity, allowing the use of only
those $k$ eigenvalues whose real part is positive, $g\sim
\exp(-{\rm Re}(k)y)\rightarrow0$ as $y\rightarrow\infty$.

In Fig.~2 we examine the BCP melt confined by one sinusoidally
patterned surface, $\sigma(x)=\sigma_q\cos(q_xx)$, with no
preference on average to one of the blocks,
$\langle\sigma\rangle=0$, for several values of surface period
$d_x=2\pi/q_x$ and for fixed value of the Flory parameter
$\chi>\chi_c$. The main effect of increasing the surface period
$d_x$ with respect to $d_0$ is to stabilize tilted lamellae, with
increasing tilt angle. Note that even for $d_x=d_0$ [Fig.~2a]
yielding no tilt, the perpendicular lamellae have a different
structure close to the surface as is induced by the surface pattern.

Although the surface interactions are assumed to be strictly local,
the connectivity of the chains causes surface-bound distortions  to
propagate into the bulk of the BCP melt. Using the analogy with the
grain boundary problem,\cite{T-A-S00} we conclude that the width
$D$ of the deformation close to the surface has a $D\sim 1/\theta$
scaling for small $\theta$, which means that
$D\sim(1-d_0/d_x)^{-1/2}$. However, in a thin film system as is
realized in experiments, this width may be too large to be detected
if $D$ is larger than the inter-surface separation $2L$.

The copolymer deformations are particularly important close to the
ODT ($\chi \gtrsim \chi_c$), 
where deviations from the optimal lamellar shape do not cost
much energy (in contrast to the strong segregation case
$\chi\gg\chi_c$), and the melt can adjust more easily to the
surface pattern. In Fig.~3 we show the same series of plots as in
Fig.~2, but with the value of the Flory parameter $\chi N=10.6$,
closer to its critical value, $\chi_c\simeq 10.49$. The perfect
lamellar shape is more distorted.

Concentration variations are governed by the profile function
$g(y)$, and increase as $\sigma_q$ increases. In Fig.~4 the Flory
parameter $\chi$ and the repeat period $d_x$ are held fixed, while
the correction $\delta\phi$ increases as $\sigma_q$ is increased
from $1$ in (a), to $2$ in (b), to $4$ in (c), in units of
$hq_0^3\phi_q$. Strong surface fields give rise to strong chain
 stretching near the surface.
Very large amplitude $\sigma_q$ increases $\delta\phi$ on the
expense of the bulk order parameter, $\phi_b$, which remains
constant. Note the similarity of these plots to the contour plots of
Ref.~\onlinecite{P-Muthu98}.

\section{BCP thin films: One homogeneous and one patterned surface}
\label{2surfaces}

Until now we have considered the semi-infinite problem of a BCP
melt confined by one patterned surface. It is of experimental and
theoretical interest to study thin films of BCP when they are
confined between a heterogeneous (patterned) surface and a second
chemically homogeneous surface. This situation is encountered when
a thin BCP is spread on a patterned surface. The second interface
is the film/air interface (neglecting any height undulations), and
is homogeneous. Usually the free
surface has a lower surface tension with one of
the two blocks. This bias can be modeled by adding a constant
$\sigma_0$ term to the $\sigma(x)$ surface field. 
For
simplicity, we assume that the $y=-L$ surface remains purely
sinusoidal while the $y=L$ surface is attractive to one of the A/B
blocks:
\begin{eqnarray}
\sigma(x)&=&\sigma_q\cos(q_xx), \qquad\qquad\mbox{at $y=-L$},\nonumber\\
\sigma(x)&=&\sigma_0, \qquad \qquad\qquad\qquad\mbox{at $y=L$.}
\end{eqnarray}
A neutral surface at $y=L$ is obtained as a special case with $\sigma_0=0$.
The striped surface pattern is at $y=-L$ and the expression 
(\ref{bulk}) for the bulk tilted phase is modified accordingly,

\begin{equation}\label{bulk2}
\phi_b=-\phi_q\cos\left[q_xx+q_y(y+L)\right]
\end{equation}
The surface free energy per unit length in the $z$-direction, Eq.
(\ref{Fs}), reads:
\begin{eqnarray}\label{Fs2}
{\cal F}_s&=&\nonumber\\
&&\int {\rm d}x
\left\{\sigma_q\cos(q_xx)\phi(x,y=-L)~+~\tau_s\phi^2(x,y=-L)\right.\nonumber\\
&&\left. +\sigma_0\phi(x,y=L)~+~\tau_s\phi^2(x,y=L)\right\},
\end{eqnarray}
where for simplicity the surface parameter $\tau_s$ is taken to be the same
on the two boundaries $y=\pm L$. 

For inter-surface separations $2L$ much larger than the bulk
lamellar period, $d_0$, the BCP morphology is similar to the
one-surface case. Bringing the two surfaces closer together changes
the melt morphology. In our formalism, $\phi_b(x,y)$, being the
zero-order approximation, remains unchanged, while the correction
$\delta \phi$ changes (recalling that the spatial variations of
$\phi_b$, Eq. (\ref{bulk2}), comes from the imposed tilt between
the bulk lamellar phase and the surface direction).
The homogeneous surface field at $y=L$ induces a
lamellar layering parallel to the surface, since the two A/B blocks
are covalently linked together. The simplest way to account for
this layering effect is to extend Eq. (\ref{dphi}) to include an
$x$-independent term $w(y)$ in our ansatz for the order parameter:
\begin{equation}\label{dphi2}
\delta\phi(x,y)=g(y)\cos(q_xx)+w(y).
\end{equation}
The free energy of the system is now written as a sum of two contributions:
\begin{equation}\label{Dgh}
\Delta {\cal F}=\int \left\{\Delta F_g+\Delta F_w\right\}{\rm d}y,
\end{equation}
where $\Delta F_g$ is given by 
\begin{eqnarray}
\Delta F_g(y)&=&-\frac12\tau\cos^2(q_yy)g^2
+\frac14h\left(q_y^2g+g^{\prime\prime}\right)^2\nonumber\\
&+&\frac12 \left(\sigma_qg+\tau_s\left(-2\phi_q g+g^2\right)\right)\delta(y+L)\nonumber\\
&+&\frac12 \tau_s\left(-2\phi_q\cos(2q_yL) g+g^2\right)\delta(y-L)
\end{eqnarray}
and 
\begin{eqnarray}\label{Dh}
\Delta F_w(y)&=&-\frac12\tau w^2
~+~\frac12h\left(q_0^2w+w^{\prime\prime}\right)^2~-~\mu w\nonumber \\
&+&\left(\sigma_0w+\tau_sw^2\right)\delta(y-L)~+~\tau_s
w^2\delta(y+L).
\end{eqnarray}

We note here that in the expansion of the free energy to second
order in $\delta\phi$ the mixed terms which couple $w$ and $g$ in
$\Delta {\cal F}$ vanish. Again, the $w$-dependent terms in the surface
free energy Eq. (\ref{Fs2}) are expressed via a Dirac delta
function. The chemical potential $\mu$ couples only to $w(y)$ and
not to $g(y)$, and fixes the total A/B ratio in the system as is
imposed by the parameter $f$ of the BCP chains (taken to be $f=1/2$ in this paper). 
The Euler-Lagrange
equation corresponding to this free energy, Eq. (\ref{Dh}), is a
linear, ordinary fourth-order differential equation. Its solution
can be written as a superposition of four exponential functions and
a constant term:
\begin{eqnarray}\label{hy}
w(y)&=&A_we^{-k_wy}+B_we^{k_wy}\nonumber\\
&+&~A^*_we^{-k^*_wy}+B^*_we^{k^*_wy}~+const.
\end{eqnarray}
with
\begin{equation}\label{k_w}
k_w^2=-q_0^2+\sqrt{\tau/h}.
\end{equation}
The boundary conditions obeyed by $w(y)$ at $y=\pm L$ are
\begin{eqnarray}
q_0^2w(\pm L)+ w^{\prime\prime}(\pm L)=0,\label{h_bcs1} \nonumber\\
2\tau_sw(-L)+ hq_0^2w^{\prime}(-L)+
hw^{\prime\prime\prime}(-L)=0,\label{h_bcs2}\nonumber \\
\sigma_0+2\tau_sw(L)-hq_0^2w^{\prime}(L)-
hw^{\prime\prime\prime}(L)=0,\label{h_bcs3}
\end{eqnarray}
and those obeyed by $g(y)$ are similar to those in Eqs. (\ref{g_bcs}):
\begin{eqnarray}
q_y^2g(\pm L)+g^{\prime\prime}(\pm L)=0,\nonumber \\
\sigma_q-2\tau_s\phi_q+2\tau_sg(-L)+hq_y^2g^{\prime}(-L)+h
g^{\prime\prime\prime}(-L)=0,\nonumber \\
-2\tau_s\phi_q\cos(2q_yL)+2\tau_sg(L)-hq_y^2g^{\prime}(L)-h
g^{\prime\prime\prime}(L)=0.
\end{eqnarray}

In general the wavenumber $k_w$ and the amplitudes $A_w$ and $B_w$
are complex. Expressions~(\ref{k_w}) and (\ref{hy}) are
identical to the zero $q$-mode response of a BCP in presence of a
surface field for a system found above its ODT temperature in the
disordered phase.\cite{epl01,mm01} Expression (\ref{hy}) has
decaying modulations, as the wavevector $k_w$ has both real and
imaginary parts. The decay length $\xi=1/{\rm Re}(k_w)$ diverges as
$(\chi-\chi_c)^{-1/2}$, while the modulation periodicity is
slightly longer than that of the bulk lamellar phase (to order
$\chi-\chi_c$). Lastly, the constant term in Eq. (\ref{hy}) [and the
chemical potential $\mu$ in Eq.~(\ref{Dh})] are determined by
requiring that the total A/B ratio of monomers is conserved, $\int
w(y){\rm d}y=0$.

Figure~5 demonstrates two different thin-film morphologies, for a
melt confined between two patterned surfaces, one at $y=-L=-2d_0$
and another which is homogeneously attractive to the B monomers at
$y=L=2d_0$. The BCP close to the lower and upper surfaces show
different behavior, as the bottom surface is chemically patterned
and the top is uniform. In Fig.~5(a) the top surface has a weak
overall preference to the B monomers (in black), 
while in (b) this attraction is stronger, thus
enhancing parallel lamellar ordering (layering).

This tilted lamellar phase confined by one homogeneous
and one patterned surface is a generalization of
the mixed (perpendicular and parallel) lamellar phase, usually
referred to as $L_M$. The latter morphology occurs when the surface
(imposed) periodicity $d_x$ is equal to the bulk periodicity $d_0$.
This ``T-junction'' morphology, shown in Fig.~6, has perpendicular
lamellae extending from the patterned surface. The homogeneous
field at the opposite surface favors a parallel orientation of the
lamellae. The crossover region between the two orientations is found
in the middle of the film, and its morphology depends on the
temperature, as can be seen by comparing 
(a) in which $\chi N=11$ with (b) where $\chi
N=10.7$. In the latter case the effect of the homogeneous field is
more evident, as parallel ordering extends from the top surface.
For homogeneous surfaces and symmetric ($f=1/2$) BCP melts, these
phases were considered unstable with respect to the $L_\parallel$
and $L_\perp$ phases, \cite{matsenJCP97,pbmm97} recalling that bulk
T-junction defects are not usually found in experiments. \cite{G-TMAC94} 
However, strong enough
modulated surface fields stabilize the tilted lamellar phases, and
in particular the $L_M$.

It is important to check the self-consistency of our free energy
variation by studying the difference between the film and bulk free
energies. We show the free energy decrease  as a function of
surface separation $2L$ in Fig.~7, where $\Delta {\cal F}$ is taken from
Eq. (\ref{Dgh}). The free energy has a characteristic oscillatory
behavior as a function of surface separation $2L$. 
The period of these energy barriers is different 
from the bulk lamellar spacing $d_0$, and depends on the tilt 
angle of the lamellae.

For separations
larger than what is shown in the figure ($2L>8d_0$) the free
energy difference tends to zero, since the correction field $\delta\phi$
has an important contribution only at a finite range away from the
surface. As one approaches the ODT temperature the free energy
difference $\Delta {\cal F}$ becomes larger compared to the bulk one
because $D$, the width of surface induced ordering, increases and
the bulk free energy decreases. Far from the ODT, i.e. in the strong
segregation limit, the bulk lamellar ordering is strong, and these
surface corrections are important only for large surface
fields $\sigma$.


\section{conclusion}\label{conclusion}

We have employed a Ginzburg-Landau expansion of the free
energy to study analytically the confinement effects of block
copolymers between two surfaces as well as the interfacial behavior
close to a single surface. For a BCP melt below its bulk ODT
temperature, tilted lamellar phases are shown to exist between one
patterned and one homogeneous surface if the 
chemical pattern is strong enough, and if no 
other defects exist inside the film.

Our approach consists of expressing the copolymer order parameter
$\phi$ as a sum of two terms. The first term accounts for the bulk
phase $\phi_b$ having the unperturbed periodicity $d_0$, while the
presumably small correction, $\delta\phi$, is entirely due to the
surface and has the surface periodicity $d_x$. \cite{epl01,mm01}
The extent of the commensurate region far from the surface is
determined by the range of the amplitude function $g(y)$ introduced
above, and it diverges at the ODT. This mean-field approach is
valid close to the ODT, but not too close, where critical
fluctuations are important.\cite{F-H87,brazovskii} In addition, the linear response field
$\delta\phi$ is assumed small, $\delta\phi\ll \phi_b$, and this can
be justified if the surface field $\sigma$ is not too large.

The deviation from the perfect diagonal lamellae $\phi_b$ relieves
the BCP elastic energy near the surface, and induces an almost
locally-parallel ordering.
 The analytic expressions presented here capture the effects
Petera and Muthukumar investigated using the self-consistent field
theory. \cite{P-Muthu98} Moreover, in the weak segregation
 limit studied here,
$\chi\approx\chi_c$, the energetic difference between the lamellar
and the disordered phases is small, leading to lamellae which are
more amenable to deformations. In this weak segregation regime the
correction to the order parameter, $\delta\phi$, gives an important
contribution to the free energy (as can be seen quantitatively in
Fig.~7), and to the resulting morphology. This description, however,
is accurate only when the surface interactions at $y=-L$ are large,
and the overall state is that of the tilted lamellae morphology.

We find a similarity between thin BCP films and that of the
symmetric tilt grain boundaries. In the latter system, the tilt
angle is determined by the externally imposed relative orientation
of the two grains. In both systems, packing frustration of the
polymer chains occur at the interface. However, the BCP film we
consider here has a richer behavior because the model contains a
real surface interacting chemically with the chains. The analogy
 shows that the width $D$ of surface deformation
scales as $D\sim(1-d_0/d_x)^{-1/2}$, for $1-d_0/d_x\ll 1$.

We consider in particular the case where one of the surfaces 
uniformly prefers on of the blocks. This can be, for example, the free surface of
the film (polymer/air interface) and is accommodated in our model
by considering an additional response field, $w(y)$. Tilted lamellae
appear at the patterned surface while parallel lamellae are created
at the opposite surface, generalizing the mixed lamellar phases
$L_M$ found for homogeneous surfaces. \cite{wang1,wang2}
 Morphologies similar to
T--junction defects are then obtained when the pattern period is the
natural period. Although the energetic penalty for a T--junction
 in the bulk is
 high, chemically interacting surfaces can stabilize these
structures.

The field $w(y)$ is also used to describe the perpendicular
$L_{\perp}$ and parallel oriented $L_{\parallel}$ lamellar phases
found in the presence of uniform surfaces. The free energies of
these phases change considerably under the influence of the surface
fields, and
 hence the features of the phase diagram may
be different from that of the strong segregation regime.\cite{T-Axx00}

\acknowledgements The authors would like to thank G. Krausch,
M. Muthukumar, G. Reiter, T. Russell, M.  Schick and U.
Steiner for useful discussions.  Partial support from the U.S.-Israel
Binational Foundation (B.S.F.) under grant No. 98-00429 and the Israel
Science Foundation founded by the Israel Academy of Sciences and
Humanities --- centers of Excellence Program is gratefully
acknowledged.

\newpage

\begin{itemize}

\item{\bf Figure~1:}
Schematic drawing of the BCP film. In (a) the BCP melt of
periodicity $d_0$ is confined by a flat patterned surface at $y=0$.
The tilt angle with the normal to the surface is $\theta$, (see Sec.
\ref{1surface}). In (b) the melt is confined by a patterned surface
at $y=-L$ and a homogeneous surface at $y=L$ (Sec. \ref{2surfaces}).

\item{\bf Figure~2:}
Tilted lamellar phase between two parallel patterned surfaces. The
surface patterning is modeled by the term $\sigma_q\cos(2\pi
x/d_x)$. The lamellae tilt angle $\theta=\arccos(d_0/d_x)$ increase
as the periodicity of the surface
  $d_x$ increase: $\theta=0$ for $d_x=d_0$ in (a), $\theta\simeq 48.1^o$
  for $d_x=\frac32 d_0$ in (b) and $\theta\approx 70.5^o$ for $d_x=3 d_0$ in (c). The correction field $\delta\phi$ is more  important for large tilt angles, as it relieves the elastic stress
  near the surface. In the plots $\sigma_q/hq_0^3\phi_q=1$ and
   the Flory
  parameter, $N\chi=11.5$.
  In this and subsequent figures $\tau_s/hq_0^3=0.1$ and maximum (A-rich) and minimum (B-rich) values
  of $\phi$ correspond to
 white and black colors, respectively.

\item{\bf Figure~3:}
The same sequence of tilt angles as in Fig.~2, but with the Flory
parameter $N\chi=10.8$ much closer to its critical value
$N\chi_c\simeq 10.49$. The sinusoidally modulated surface field
$\sigma(x)$ causes  strong distortions of the tilted lamellar phase
[parts (b) and (c)], even farther away from the surface.

\item{\bf Figure~4:}
Variation of the tilted lamellar morphology with
$\sigma_q$, the amplitude
of surface fields. The value of $\sigma_q/hq_0^3\phi_q$ is varied
from $1$ in (a), to $2$ in (b), to $4$ in (c), while the
Flory parameter $N\chi=11>N\chi_c$, and the repeat period
$d_x=\frac32 d_0$ are held fixed.

\item{\bf Figure~5:}
A contour plot for a BCP melt between a top homogeneous surface at
$y=L$, and a modulated bottom surface at $y=-L$. The inter-surface
separation $2L$ is chosen to be $4d_0$ and the modulated surface
amplitude is $\sigma_q/hq_0^3\phi_q=2$. The homogeneous top field
is $\sigma_0/hq_0^3\phi_q=0.6$ in (a), and larger
$\sigma_0/hq_0^3\phi_q=2$ in (b), preferring adsorption of the B
monomers (black). The effect of this field is clearly seen close to
the $y=L$ surface. In both plots the Flory parameter was chosen to
be $N\chi=11$ and the repeat period of the bottom surface is
$d=\frac32 d_0$.

\item{\bf Figure~6:}
A BCP confined film showing a crossover from perpendicular lamellae
at the $y=-L=-2d_0$ surface to parallel lamellae at $y=L$. The
pattern on the bottom surface, $\sigma(x)=\sigma_q\cos(q_0x)$, has the bulk periodicity $d_0$, and amplitude $\sigma_q=2hq_0^3$, while the
top surface ($y=L$) is homogeneously attractive to the B polymer
(in black), $\sigma_0=4hq_0^3$. In (a) the Flory parameter is
$N\chi=11$, while in (b) the temperature is closer to the ODT,
$N\chi=10.7$.

\item{\bf Figure~7:}
The correction free energy $\Delta {\cal F}$ (divided by ${\cal F}[\phi_b]$, the
bulk lamellar free energy) as a function of the film thickness
$2L$. Lengths are scaled by the bulk lamellar period $d_0$. The period of the oscillations characterizing the 
free energy $\Delta {\cal F}$ is different
from $d_0$, because the lamellae are tilted with respect to the surface. For all points in the plot, the modulated surface
field at $y=-L$ has an amplitude $\sigma_q/hq_0^3\phi_q=0.5$ and a
period $d_x=2\pi/q_x=\frac32 d_0$, while the surface field at $y=L$ is
homogeneous with $\sigma_0=\sigma_q$. The Flory parameter is
$N\chi=10.7$ and $\tau_s=0.7hq_0^3$.

\end{itemize}
\end{multicols}

\begin{figure}[b]
\epsfig{file=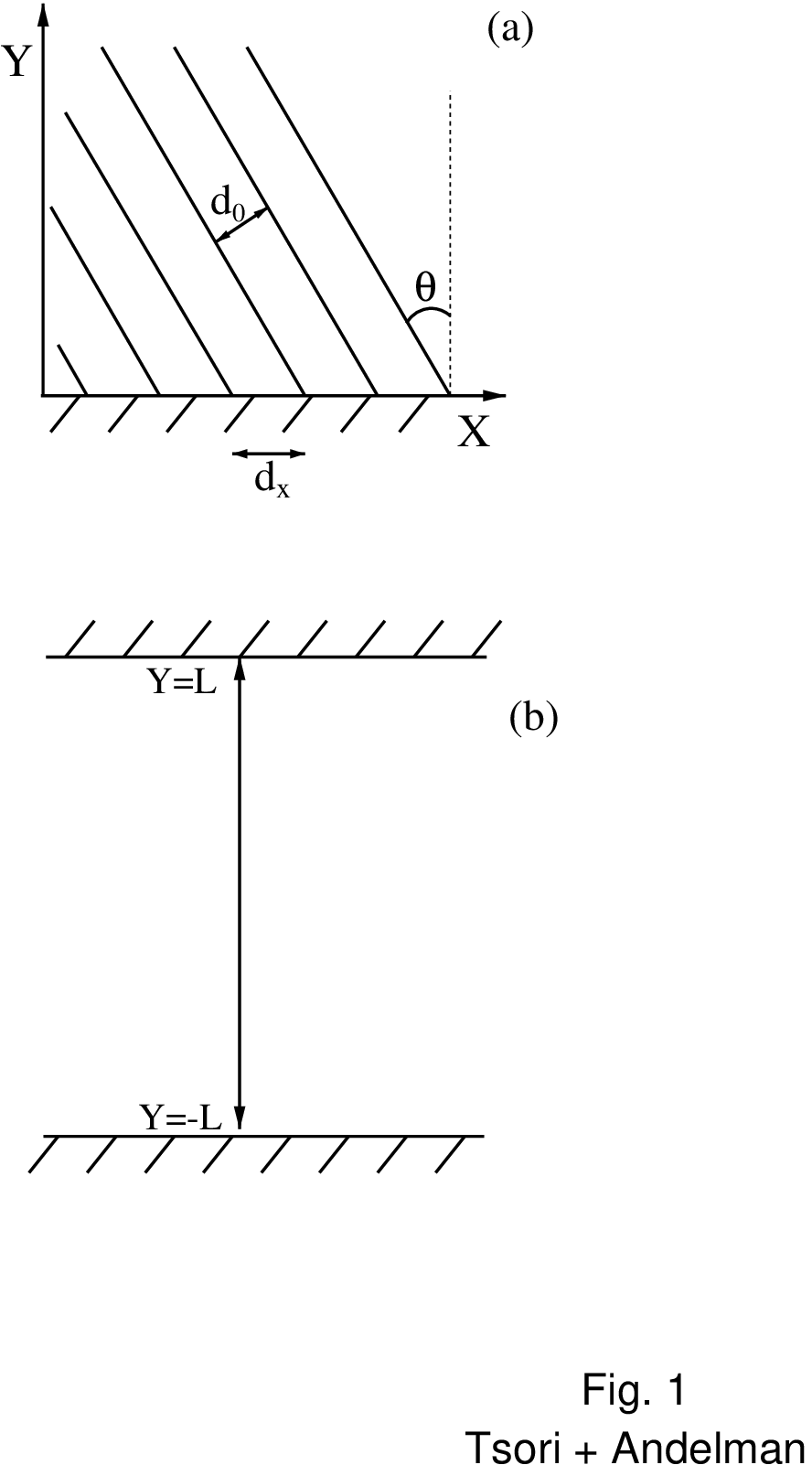,scale=0.6}
\end{figure}
\vspace{1cm}

\begin{figure}[b]
\epsfig{file=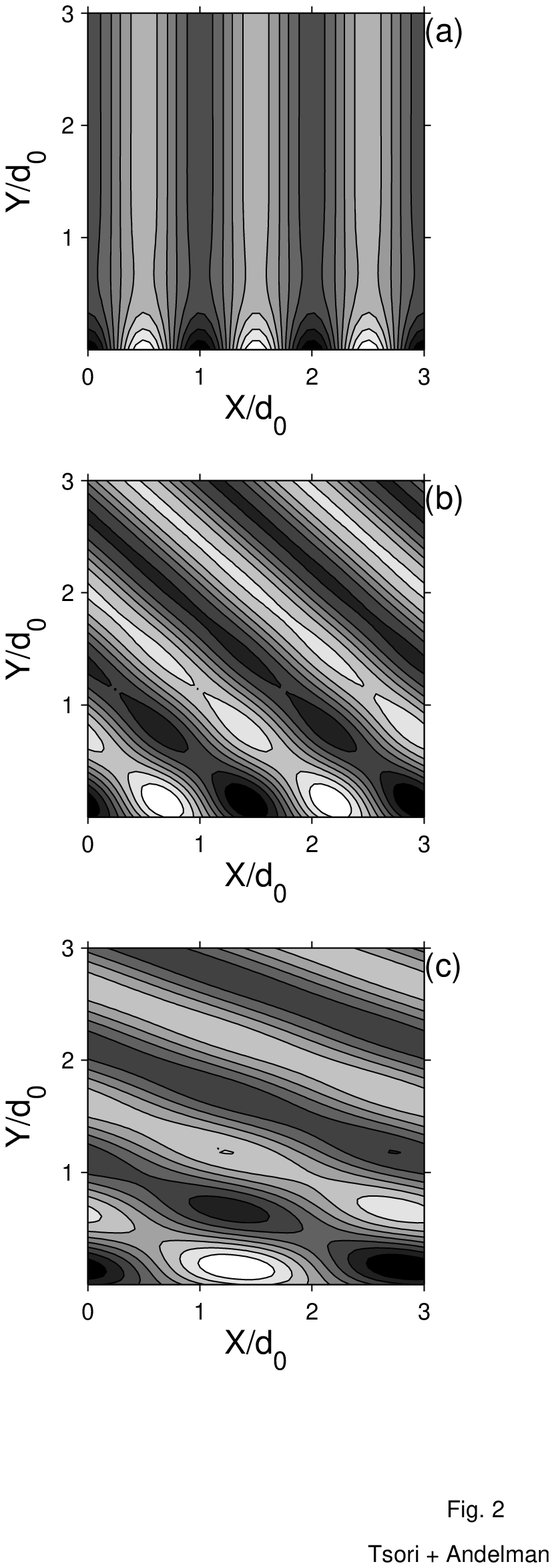,scale=0.95}
\end{figure}
\vspace{1cm}

\begin{figure}[b]
\epsfig{file=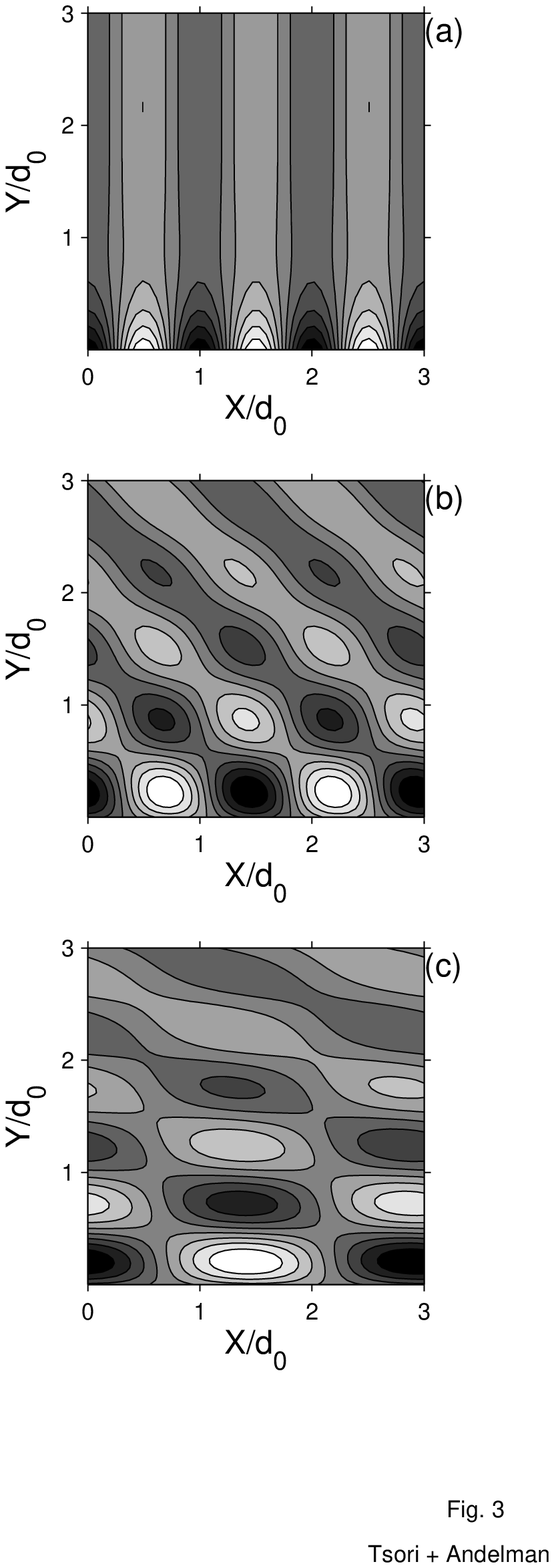,scale=0.95}
\end{figure}
\vspace{1cm}

\begin{figure}[b]
\epsfig{file=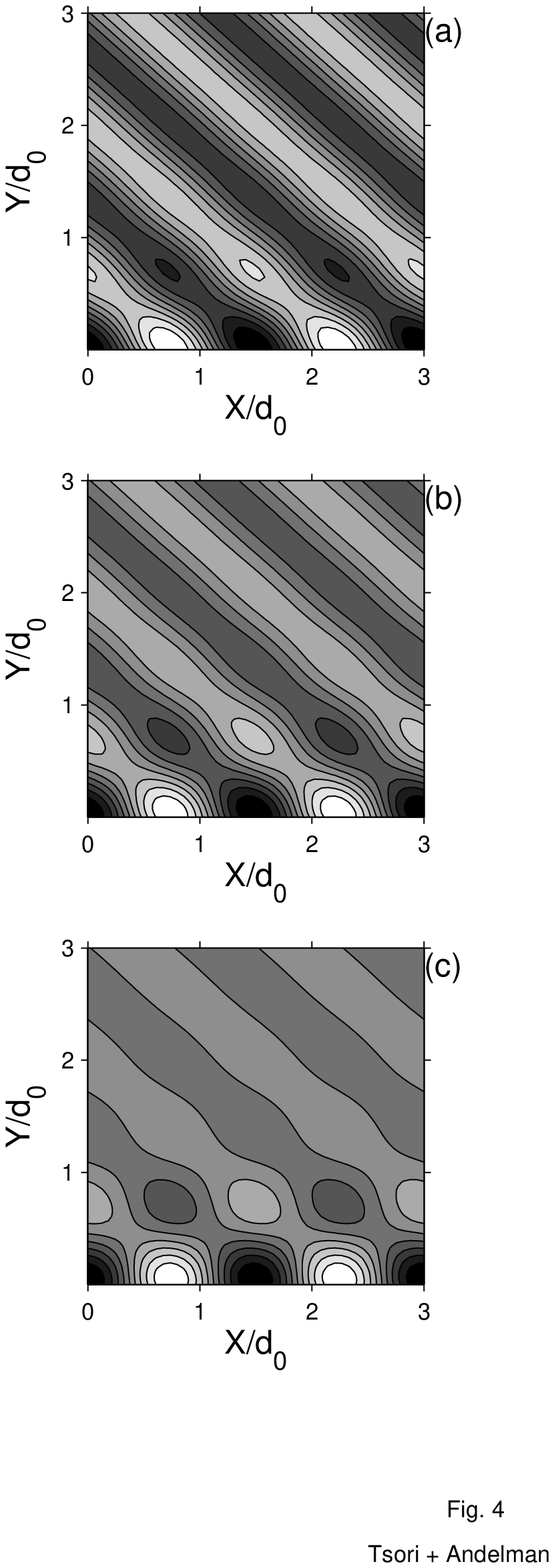,scale=0.95}
\end{figure}
\vspace{1cm}

\begin{figure}[b]
\epsfig{file=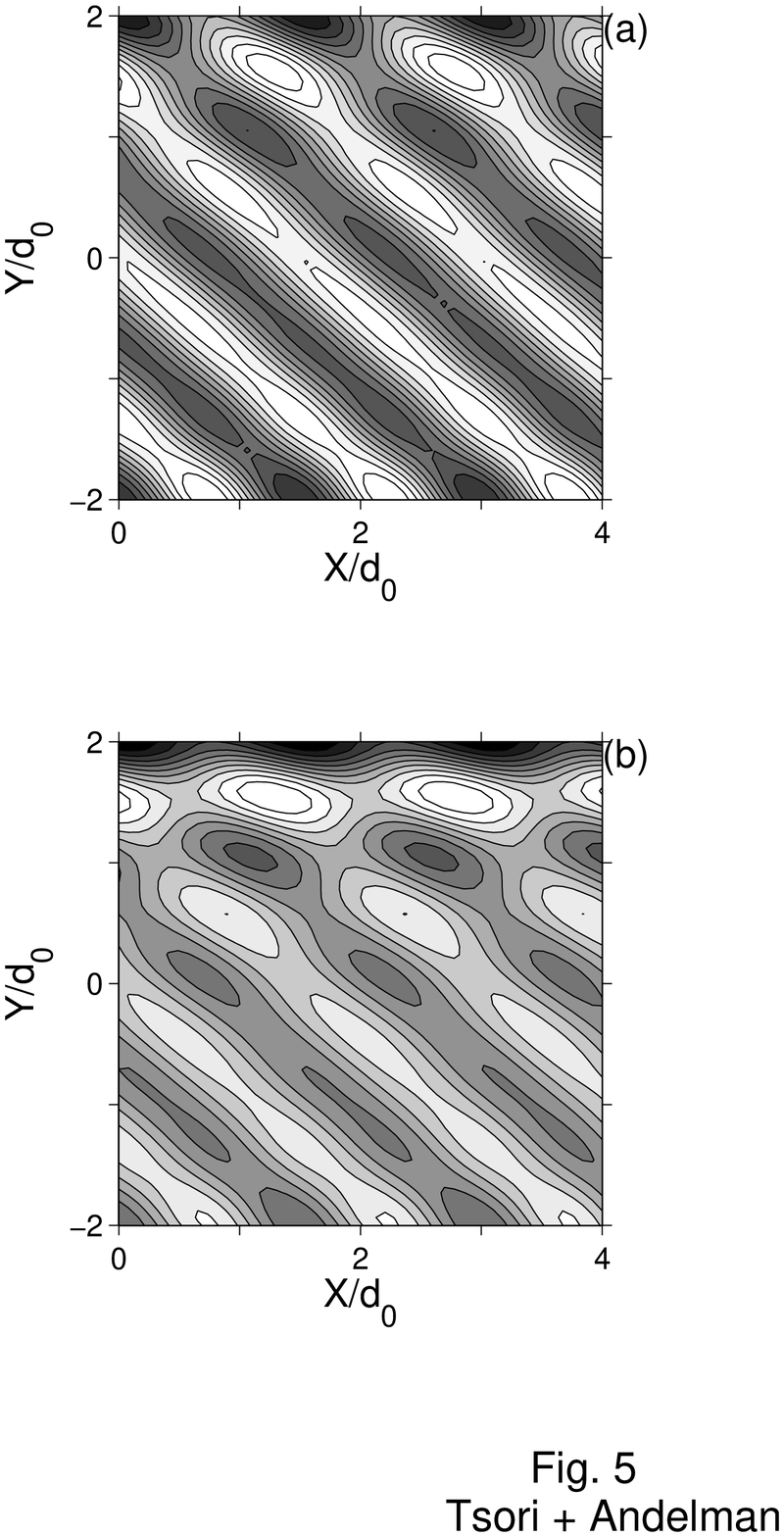,scale=0.65}
\end{figure}
\vspace{1cm}

\begin{figure}[b]
\epsfig{file=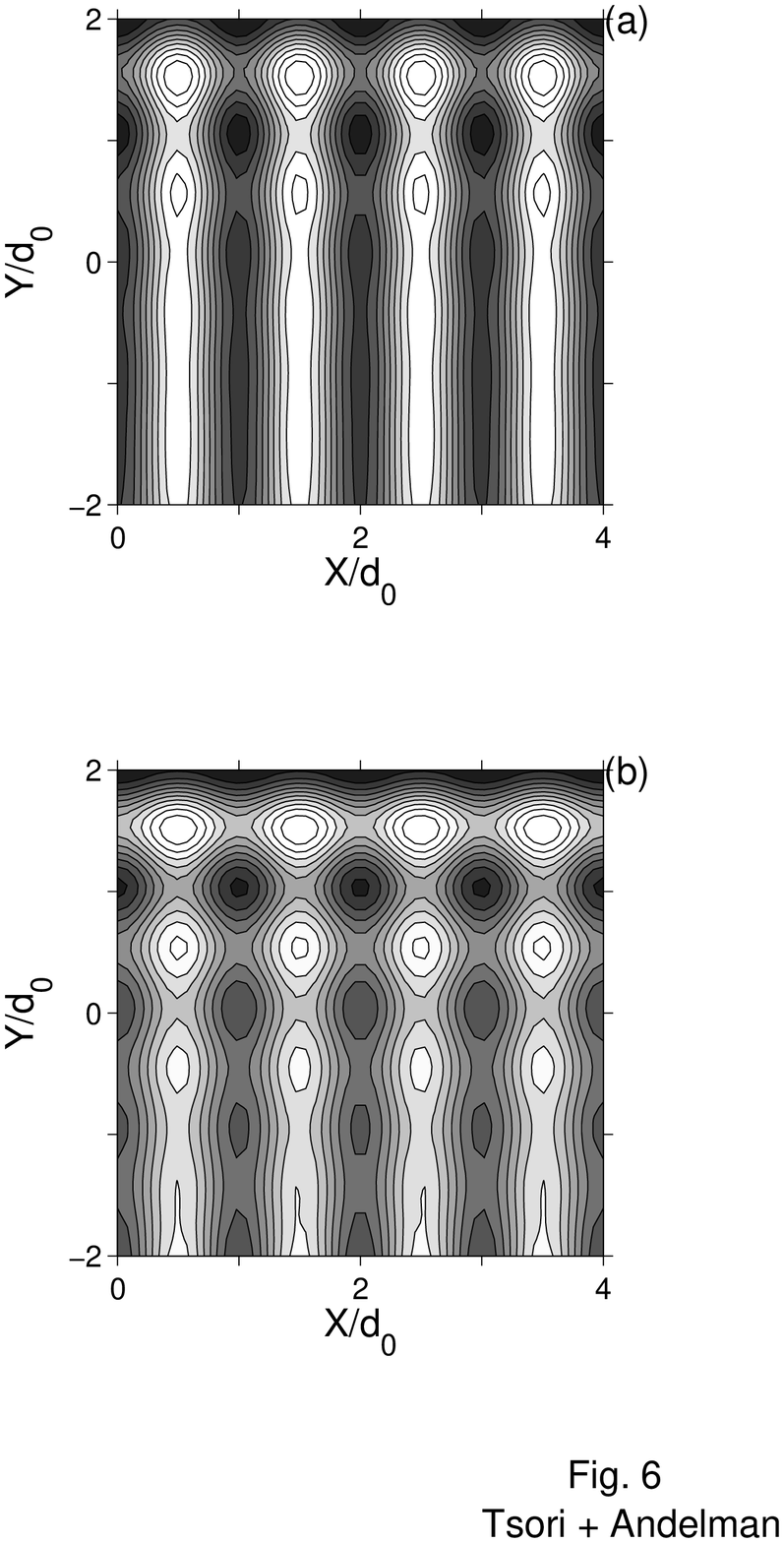,scale=0.65}
\end{figure}
\vspace{1cm}

\begin{figure}[b]
\epsfig{file=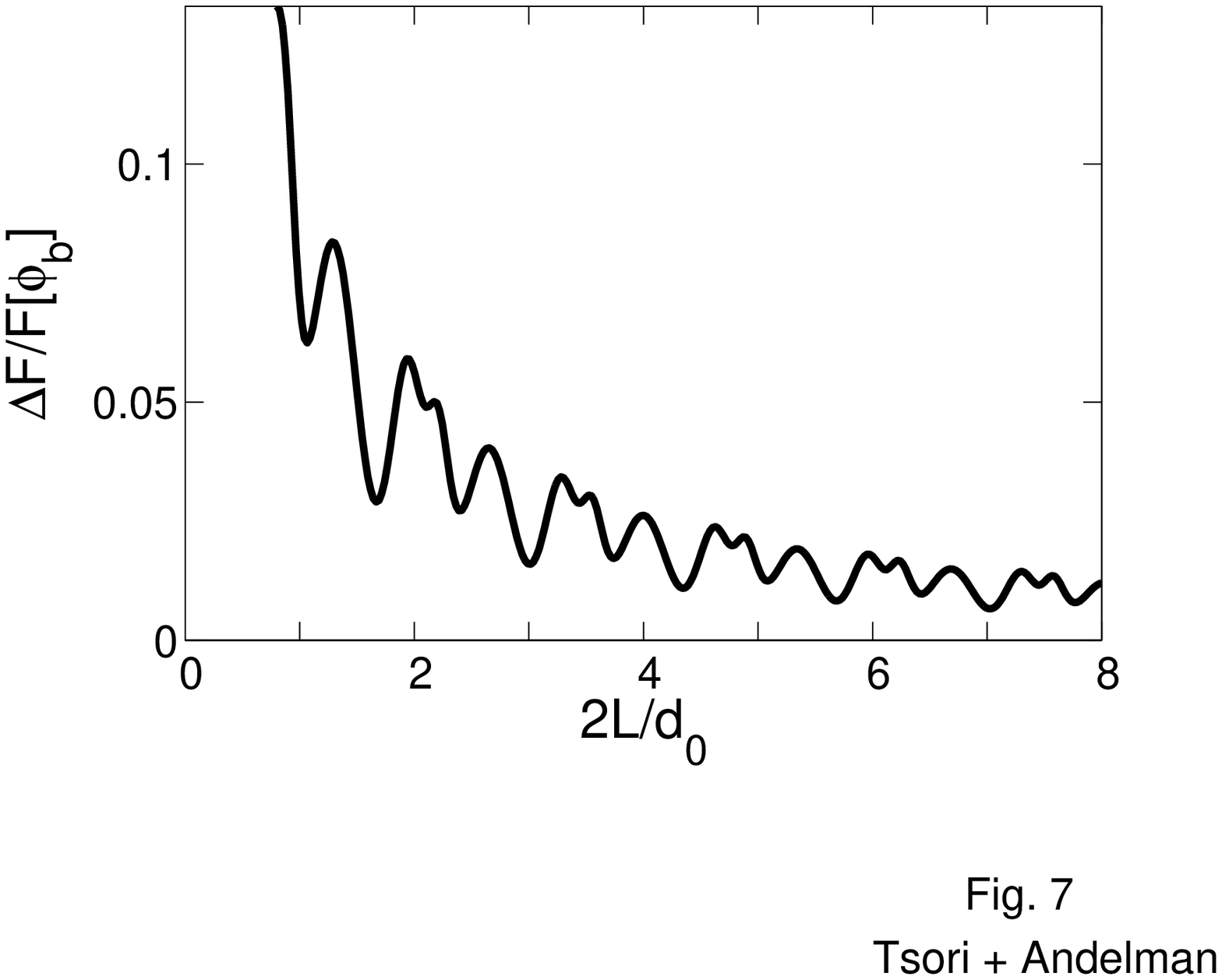,scale=0.5}
\end{figure}
\vspace{1cm}

\end{document}